\documentstyle[12pt]{article}

\catcode `\@=11
\@addtoreset{equation}{section}

\catcode `\@=12

\newcommand{\be}{\begin{equation}}
\newcommand{\en}{\end{equation}}
\newcommand{\bea}{\begin{eqnarray}}
\newcommand{\ena}{\end{eqnarray}}
\newcommand{\beano}{\begin{eqnarray*}}
\newcommand{\enano}{\end{eqnarray*}}
\newcommand{\bee}{\begin{enumerate}}
\newcommand{\ene}{\end{enumerate}}

\begin{document}

\thispagestyle{empty}
 
\vspace*{1cm}

\begin{center}
{\Large \bf Multiplication of Distributions in one dimension: possible
approaches and applications to $\delta$-function and its derivatives}  
\vspace{2cm}\\

{\large F. Bagarello}
\vspace{3mm}\\
  Dipartimento di Matematica ed Applicazioni, 
Fac.Ingegneria, Universit\`a di Palermo, I - 90128  Palermo, Italy\\
E-mail: Bagarello@Ipamat.math.unipa.it
\vspace{2mm}\\
\end{center}

\vspace*{2cm}

\begin{abstract}
\noindent 
	We introduce a new class of multiplications of distributions in one dimension 
merging together two different regularizations of distributions. Some of the features
of these multiplications are discussed in a certain detail.

 We use our theory to study a certain number of examples, involving
products between Dirac delta functions and its successive derivatives. 

\end{abstract}

\vfill

\newpage

\section{Introduction}

In this paper we propose a definition of a new class of multiplication of
distributions. The reason for such a new definition essentially relies in
the possibility of extending the usual product of functions to the product
of two delta functions centered at the same point, together with their
derivatives. 

The usual way in which a
multiplication of two distributions, $T_1$ and $T_2$, is defined can be
summarized in three steps: first, one regularizes these distributions using
some $'$trick$'$, in order to obtain continuous (or even more regular)
functions $T_1^{(r)}$ and $T_2^{(r)}$; second, $T_1^{(r)}$ and $T_2^{(r)}$ are
multiplied (in the sense of distributions). Finally, one tries to recover a
result using the same limiting procedure which returns $T$ from
the function $T^{(r)}$. 

In the literature plenty of methods for regularizing distributions have been
proposed, \cite{breme}-\cite{col}. In the following we will discuss essentially 
two of these methods, which are the main ingredients in the definition of our
multiplication.

The first method consists in the analytic
continuation of a distribution, first proposed in \cite{breme} and then used
by Li Bang-He, \cite{li} and \cite{li2}, in the framework of Non-standard
analysis. For reader's convenience we recall here and in Section 2 the basic
definitions and results on this method.

Given a distribution $T$ with compact support, in \cite{breme} the authors
define a function
$$
{\bf T}^0(z) \equiv (1/2\pi i)\, T\cdot (x-z)^{-1}
$$
which they prove to be holomorphic in $z$ in the whole $z$-plane minus the
support of $T$. They further extend the class for which the above definition
makes sense in order to include also distributions which do not have
compact support. In particular they are also able to compute the  analytic
continuation of distributions like $(x+i\epsilon)^{-1}$ and $P(x^{-n})$.
Moreover, they also discuss the possibility of recovering $T$ by taking a
suitable limit for $\epsilon \rightarrow 0$ of the following function 
\be
T_{red}(x,\epsilon) \equiv {\bf T}^0(x+i\epsilon)-{\bf T}^0(x-i\epsilon).
\label{red}
\en

In \cite{breme}, analytic continuation is used to define different
multiplications  of distributions. In particular, given two distributions $S$
and $T$ for which the analytic continuation makes sense, the authors define a
multiplication $(S\odot T)$ as  
$$
(S\odot T)(\phi )\equiv \lim_{\epsilon \rightarrow 0} \int
_{-\infty}^{\infty} S_{red}(x,\epsilon) 
T_{red}(x,\epsilon) \, \phi(x) \, dx,
$$
whenever this limit exists for any test function $\phi \in {\cal D}(K)$,
where $K\subseteq {\bf R}$ is the support of the test functions. (From now on
we will use simply ${\cal D}$ instead of ${\cal D}(K)$).

In particular, the authors prove that, if $S(x)$ and $T(x)$ are continuous
functions, then  $S_{red}(x,\epsilon)$ and $T_{red}(x,\epsilon)$ converge
respectively to $S(x)$ and to $T(x)$ uniformly. Consequently, under this
hypothesis, the above limit exists and it is equal to
$\int_{-\infty}^{\infty}S(x)\, T(x) \, \phi (x) \, dx$. Thus, for continuous
functions, this multiplication reduces to the ordinary product of functions.
If $S(x)$ and $T(x)$ are arbitrary distributions, then the limit may or may
not exist.
\vspace{4mm}

In \cite{miku}, \cite{fish}, \cite{col} and \cite{zhi} it is discussed a
different approach to extract a "regular" part from a given distribution.
This method, called $'$of the sequential completion$'$, makes use of the so
called  $\delta$-sequences to regularize the distributions.
In particular, one uses
 the well known property of the distributions belonging to the dual of ${\cal
D}$, ${\cal
D'}$, of returning $C^{\infty}-$functions after that their convolution
with functions in ${\cal D}$ is taken. We start by considering a function
$\phi \in {\cal D}$ with support in $[-1,1]$ and such that
$\int_{-\infty}^{\infty} \phi (t) \, dt\,=1$. With such a $\phi$ we can define
a so called $\delta$-sequence by $\delta_n(x) \equiv n \phi (nx)$, see
\cite{col}. Obviously, given any distribution $T \in {\cal D'},$
the convolution $(T*\delta_n)(x)$ is a $ C^{\infty}-$ function, for any fixed
$n$. Furthermore its limit in ${\cal D'}$ is exactly $T$. For this reason
$\delta_n(x)$ can be thought of as an approximate identity.

In \cite{col}, Chapter 2, it is sketched how to use the above 
property of the convolution to define a possible multiplication. Let us start
again with two distributions $S$ and $T$ in ${\cal D'}({\bf R}^m)$. Let
$\delta_n({\bf x})$ be a generic $\delta$-sequence, then $(T*\delta_n)({\bf
x})$ and $(S*\delta_n)({\bf x})$ are $ C^{\infty} -$ functions on ${\bf R}^m$
for any fixed $n$. One says that $T$ and $S$ are multipliable if, for any
$\delta -$sequence, the product $(T*\delta_n) \cdot (S*\delta_n)$ converges in
${\cal D'}({\bf R}^m)$ to a limit independent of $(\delta_n)$, when $n
\rightarrow \infty $. In particular in \cite{col} it is shown that this definition
does \underline{not} allow the computation of $\delta^2$. Moreover it is also
 mentioned that, for continuous functions, the above definition
coincides with the usual multiplication. 

The possibility of defining a new multiplication ,$\otimes$ (not to be
confused with the tensor product), which generalizes in some sense both
definitions above, may have a certain relevance if it allows the computation of
the product of "more" or "more interesting" distributions.

Along this paper we restrict our interest to one spatial dimension. At a first sight,
this may seem a strong  physical limitation, but, in our opinion, this is not true.
In fact, we know since Wightman, \cite{wig}, that in Relativistic Quantum Field
Theory the expectation values of the fields are distributions in $S'({\bf R}^m)$,
$\forall m\geq1$, and that the fields themselves are operator-valued distributions.
Therefore, independing of the space dimension, a special care must be used in order
to define  products of fields which appear, for instance, in the definition of the
density of the Lagrangian. This problem is discussed in many details in \cite{col},
where it is also pointed out the link between a correct definition of these products
and the disappearance of the divergences of the theory. Like in \cite{col}, the final
aim of our work should be to discuss some physical relevant theory in $3+1$
dimensions, like QCD.  However, there
exist many interesting relativistic models already in $1+1$ which can be used to
discuss the utility of our method in Quantum Field Theory, like, for instance,
the Schwinger model, \cite{franco}.

Finally, it is worthwhile to notice that an algebraic approach for the
extension of the multiplication of distributions could be set up using a quite
natural structure, that is the one given by partial*-algebras, see \cite{tra}.
However, this is not the line we will follow in this paper.
 
The paper is organized as follows:

in the following Section we introduce the definition of our multiplication
and we discuss some of its properties;

in Section 3 we give some examples of products which can be
computed using our definition. 

in Section 4 we summarize and comment the results.

We end the paper with an Appendix on some applications to quantum mechanics of
the results obtained in Section 3.

\section{Definition of the multiplication}

We start this Section recalling some known results concerning the products
discussed in the Introduction, see \cite{breme,col}.

 In \cite{breme} the authors prove the following result:
\vspace{4mm}

\noindent
	{\bf Theorem 1}.--
 Given a distribution $T$ with compact support the function 
\be
\label{analitic} 
 {\bf T}^0(z) \equiv \frac{1}{2\pi i}\, T\cdot(x-z)^{-1} 
\en
exists and is holomorphic in $z$ in the whole $z$-plane minus the support of
$T$. 

If $T(x)$ is a continuous function with compact support, then 
$T_{red}(x,\epsilon) $ converges uniformly to $T(x)$ on the whole real axis 
for $\epsilon \rightarrow 0^+$. 

If $T$ is a distribution in ${\cal D'}$ with compact support then
$T_{red}(x,\epsilon)$ converges to $T$ in the following sense 
$$
T (\phi) = \lim_{\epsilon \rightarrow 0} \int_{-\infty}^{\infty}
T_{red}(x,\epsilon) \, \phi (x) \, dx
$$
for every test function $\phi \in {\cal D}$.\hfill $\Box$ 
\vspace{4mm}

As already mentioned in the Introduction it is possible to give a meaning to
 definition (\ref{analitic}) even for other distributions. The authors define
the space ${\cal V}$ as the subspace of all the functions in $C^{\infty}$ with
arbitrary support, ${\cal E}$, with the following properties:

$i) \hspace{1.1cm} \phi (x)\; |x| \leq k_0 \mbox{   for} \; |x| \rightarrow
\infty, $

$
ii) \hspace{1cm}  \phi^{(n)}(x)\; |x| \leq k_n \mbox{  for} \; x \rightarrow
\infty, $

where $k_0, k_1,...$ are constants. The convergence is defined as in ${\cal
E}$.

Denoting with ${\cal V'}$ the dual space of ${\cal V}$, the authors prove that 
theorem 1 can be stated even for distributions in ${\cal V'}$, so to include
in their work also distributions with a slow fall off like, for instance,
$(x+i\epsilon)^{-1}$ and $P(x^{-n})$.

In \cite{breme} and in \cite{li} some examples of analytic representation of
distributions are computed. We report here only the ones that we will use in
the following. 
 \bea
T(x) = \delta (x) & \Rightarrow & T_{red}(x,\epsilon) = \frac{\epsilon}{\pi
(x^2+\epsilon^2)};\\
T(x) = \delta' (x)& \Rightarrow & T_{red}(x,\epsilon) = \frac{-2}{\pi}\,
\frac{x\, \epsilon}{(x^2+\epsilon^2)^2};\\
T(x) = \delta'' (x) & \Rightarrow & T_{red}(x,\epsilon) =
\frac{2}{\pi}\, \frac{3x^2\epsilon-\epsilon^3}{(x^2+\epsilon^2)^3}.
\ena
where $T_{red}(x,\epsilon)$ has been defined in (\ref{red}).

\vspace{5mm}
The main informations relative to the method of sequential completion can
be found in \cite{col} and they follow essentially from a very well known
result on the regularity of the convolution of distributions and test
functions. Stated as a single  theorem we have
\vspace{4mm}

\noindent
	{\bf Theorem 2}.--
Let $\phi \in {\cal D}({\bf R})$ be a given function with supp $\phi \subseteq
[-1,1]$ and $\int \phi (x) \, dx =1$. We call $\delta-$sequence the sequence
$\delta_n,\, n\in {\bf N},$ defined by $\delta_n(x) \equiv n\, \phi(nx)$.

 Then, $\forall \,  T \in {\cal D'}({\bf R})$ the convolution $T_n \equiv
T*\delta_n$ is a $C^{\infty}-$function, for any fixed $n\in {\bf N}$. This
sequence converges to $T$ in the topology of ${\cal D'}$, when $n \rightarrow
\infty$.

Moreover, if $T(x)$ is a continuous function with compact support then $T_n$
converges uniformly to $T(x)$.\hfill $\Box$ 
\vspace{4mm}

So far, we have summarized known results which will be useful in the
following.
We are now ready to define our multiplication. Since we will use both
regularizations above  we will be able to define this multiplication only for
those distributions for which both the analytic continuation and the
convolution with a $\delta-$sequence exist. From the previous discussion it is
clear that the stronger requirement is the existence of the analytic
continuation, which is ensured only for distributions with a certain decay at
infinity, that is, for distributions in  ${\cal V'}$. On the contrary, it is well
known that it is always possible to construct the convolution of a distribution in
${\cal D'}$ with any $\delta-$sequence. Therefore we will be able to
define our multiplication only for distributions in ${\cal V'}$. This is not
an heavy constraint since all the derivatives of a delta distribution belong
to this class.

For any couple of distributions $T,S \, \in
{\cal V'}, \, \forall \, \alpha, \beta >0$ and $\forall \, \Psi \, \in {\cal
D}$ we define the following quantity: \be (S\otimes T)_n^{(\alpha,\beta)}(\Psi
) \equiv \frac{1}{2} \int_{-\infty}^{\infty} [S_n^{(\beta)}(x)\, 
T_{red}(x,\frac{1}{n^\alpha}) + T_n^{(\beta)}(x)\,
S_{red}(x,\frac{1}{n^\alpha})]\, \Psi (x) \, dx  \en 
where
\be
S_n^{(\beta)}(x) \equiv (S*\delta_n^{(\beta)})(x),
\label{conv}
\en
with $\delta_n^{(\beta)}(x) \equiv n^{\beta} \Phi (n^{\beta}x)$.

It is worthwhile to notice that $\delta_n^{(\beta)}$ is a $\delta$-sequence
$\forall \beta >0$, since it satisfies all the requirements discussed in
\cite{col}.

It is easy to see that, for any choice of $\alpha, \beta,\, T,\, S$ and
$\Psi$, $(S\otimes T)_n^{(\alpha,\beta)}(\Psi )$ is well defined. What may or
may not exist is its limit, when $n$ diverges.  \vspace{4mm}

\noindent
	{\bf Definition 1}.--
Given two distributions $S$ and $T$ in ${\cal V'}$ for which the above limit exists,
we define $(S\otimes T)_{(\alpha,\beta)}(\Psi )$ as:
\be
\label{def}
(S\otimes T)_{(\alpha,\beta)} (\Psi )\equiv \lim_{n \rightarrow \infty}
(S\otimes T)_n^{(\alpha,\beta)}(\Psi ) \en
\vspace{4mm}

{\bf Remark}--
We  want to stress that the definition (\ref{def}) really defines infinitely
many multiplications of distributions. In order to obtain {\underline {one
definite}} product we have to fix the positive values of $\alpha$ and $\beta$
and the particular function $\Phi$ which is used to construct the
$\delta$-sequence. In the next Section it will appear clear that different
choices of $\alpha, \beta$ and $\Phi$ generate inequivalent products and that
a clever choice is crucial for getting a "larger" set of distributions for
which the multiplication  in (\ref{def}) exists.
 \vspace{5mm}

We see that the multiplication defined in (\ref{def}) makes reference to both
the regularizations discussed above. The main result we want to discuss
here is that, like for the products in \cite{li} and \cite{col}, the 
multiplication $\otimes_{(\alpha,\beta)}$ extends the usual multiplication 
of continuous functions for all positive $\alpha$ and $\beta$. We can
prove, in fact, the following \vspace{4mm}

\noindent
	{\bf Proposition 3}.--
Let $T(x)$ and $S(x)$ be two continuous functions with compact supports, and
$\alpha$ and $\beta$ any couple of positive real numbers. Then:

$i) \;T_n^{(\beta)}(x)\cdot S_{red}(x,\frac{1}{n^\alpha}) \mbox{ converges
uniformly to } S(x)\cdot T(x);$

$ii) \; \forall \, \Psi \in D \Rightarrow (T\otimes S)_{(\alpha,\beta)}(\Psi) =
\int_{-\infty}^{\infty} T(x)\, S(x) \, \Psi (x)\, dx$
\vspace{5mm}

\noindent
	{\underline {Proof}}

The first statement is a simple consequence of the uniform convergence of the
functions $T_n^{(\beta)}(x)$ and $S_{red}(x,\frac{1}{n^\alpha})$ to the
continuous functions $T(x)$ and $S(x)$ respectively, both with compact support.
One can easily check that
$$
| T_n^{(\beta)}(x)\, S_{red}(x,\frac{1}{n^\alpha}) - T(x)\, S(x)| < \epsilon
$$
for any $n$ bigger than a certain $n_0$ depending only on $\epsilon$ and
not on $x$.

The second statement easily follows from the first.
\hfill $\Box$
\vspace{5mm}

It is furthermore very easy to see that the product $(S\otimes T)_
{(\alpha,\beta)}$ is a linear functional on ${\cal D}$ due to the linearity of the
integral and to the properties of the limit. The continuity of such a
functional is, on the contrary, not obvious at all, since it was not ensured
already for the product defined in \cite{li}. (In the next Section, however,
all the examples discussed will appear to be continuous).

\section{Examples}

In this Section we are going to show that, fixing with care the function
$\Phi, \alpha$ and $\beta$, the multiplication defined in (\ref{def})
is very powerful and can be used to extend reasonably the known
multiplications. We begin this Section by giving the expressions of the
regularized distributions we will use in the examples. From formulae (2.2),
(2.3) and (2.4) we easily deduce
 \bea
\delta_{red}(x,\frac{1}{n^\alpha}) & = &\frac{1}{\pi n^\alpha}\frac{1}
{(x^2+\frac{1}{n^{2\alpha}})};\\
\delta'_{red}(x,\frac{1}{n^\alpha})  & = &
\frac{-2}{\pi n^\alpha}\,\frac{x}{(x^2+\frac{1}{n^{2\alpha}})^2};\\
\delta''_{red}(x,\frac{1}{n^\alpha}) & = &
\frac{2}{\pi n^\alpha}\, \frac{3x^2-\frac{1}{n^{2\alpha}}}{(x^2+\frac{1}
{n^{2\alpha}})^3}.
\ena
\vspace{3mm}

From the definition (\ref{conv}) we obtain
\bea
\delta_n^{(\beta)}(x) & = & n^\beta \Phi(n^\beta x);\\
\delta_n^{'(\beta)}(x) & = & n^{2 \beta} \Phi_1(n^\beta x);\\
\delta_n^{''(\beta)}(x) & = &  n^{3 \beta} \Phi_2(n^\beta x).
\ena
where $\Phi_1(x) \equiv \frac{d\, \Phi}{dx}(x)$ and $\Phi_2(x) \equiv
\frac{d\, \Phi_1}{dx}(x)$.

\vspace{3mm}

We are now ready to discuss the examples.

\vspace{4mm}

{\bf Example 1: $(\delta \otimes \delta)_{(\alpha,\beta)}$}

The product $(.\otimes .)_{(\alpha,\beta)}$ of the distributions $\delta (x)$
with itself is defined, by (\ref{def}), as follows:

For any $\Psi \in {\cal D}$ we have
$$
(\delta \otimes \delta )_{(\alpha,\beta)}(\Psi ) \equiv \lim_{n \rightarrow
\infty} \int_{-\infty}^{\infty} \delta_n^{(\beta)}(x)\,
\delta_{red}(x,\frac{1}{n^\alpha})\, \Psi (x) \, dx
$$
Using formulas (3.1) and (3.4), and changing the integration variable,
$t=n^{\beta}x$, we get 
\be
 (\delta \otimes \delta )_{(\alpha,\beta)}(\Psi ) \equiv \lim_{n
\rightarrow \infty} \frac{1}{\pi n^{\alpha-2\beta}} \int_{-1}^{1}
\frac{\phi (t) \Psi (t/n^{\beta}) \, dt}{t^2+1/n^{2(\alpha-\beta)}}
\label{delta2}
\en

\noindent
where also the property of the support of $\Phi$ has been used to restrict the
integration limits.

To compute this integral, as well as all the others that will appear in the
following examples, we will use Lebesgue Dominated Convergence Theorem (LDCT).
 
The sequence $f_n(t)$ whose convergence and boundness we have to
discuss is 
$$\frac{1}{\pi
n^{\alpha-2\beta}}  \frac{\phi (t) \Psi (t/n^{\beta})}
{t^2+1/n^{2(\alpha-\beta)}}.
$$
 It is easily seen that there exists a function $g(t)$ such that $|f_n(t)|
\leq |g(t)|$ a.e. in $[-1,1]$, as it is required by the LDCT, if
$\alpha-2\beta \geq 0$ and if  we require to $\Phi$ to be of the
form   
\be
    \Phi(x)  = \left\{
          \begin{array}{ll}
            \frac{x^m}{F} \cdot \exp\{\frac{1}{x^2-1}\},    &       |x| <1 \\
           0,   &       |x| \geq 1. 
       \end{array}
        \right. 
\label{fi}
\en
where $m$ is a natural number and $F$ is a normalization constant
which gives $\int_{-1}^{1} \Phi(x)\, dx=1$. It is worthwhile to notice
that, in order not to have $\int_{-1}^{1} \Phi(x)\, dx=0$, $m$ must be  even,
to prevent $\Phi$ from being an odd function.

One can see that, if $m>1$, then the required function $g(t)$ has the form
$g(t)=\frac{ML}{\pi F}|t|^{m-2}$, where $M\equiv \sup_{t\in
]-1,1[}\exp\{\frac{1} {x^2-1}\}$ and $L\equiv \sup_{t\in ]-1,1[}|\Psi(t)|$. Of
course $g(t)$ is integrable in $[-1,1]$.
\vspace{3mm}
Now, depending on the values of $\alpha$ and $\beta$, we can define different
products:

i) if $\alpha=2\beta$  one easily proves that the limit of
the sequence $f_n(t)$ is the function $f(t)= \frac{\Phi(t) \Psi(0)}{\pi t^2}$.
This convergence is punctual in $]-1,1[$ and therefore it implies the
convergence almost everywhere in this interval.

ii) if $\alpha>2\beta$ the sequence $\{f_n(t)\}$ still converges but its limit
is now $f(t)=0$, due to the presence of the decreasing factor
$\frac{1}{n^{\alpha-2\beta}}$ in the definition of $f_n(t)$.

Defining the following quantities
\be
A_j \equiv \int_{-\infty}^{\infty} \frac{\Phi(t)}{t^j}\, dt
\label{aj}
\en
whenever they exist, and using the LDCT we
get 
\be
    (\delta \otimes \delta)_{(\alpha,\beta)}(\Psi)  = \left\{
          \begin{array}{ll}
          \frac{1}{\pi}A_2 \delta(\Psi),    &       \alpha=2\beta \\
           0,   &       \alpha>2\beta. 
       \end{array}
        \right. 
\label{res1}
\en

\vspace{4mm}
{\bf Remarks--} 
(a) The above result, for $\alpha>2\beta$, numerically coincides with the one
given by the neutrix product discussed by Zhi and Fisher, see \cite{zhi}.

(b) $A_2$ exists surely whenever we take $\Phi$ as in (\ref{fi}) with $m>1$. In
order to get well-defined $A_j$, with $j>2$, we will be led to consider bigger
values of $m$.

\vspace{4mm}
{\bf Example 2: $(\delta \otimes \delta')_{(\alpha,\beta)}$}

The above multiplication is defined by the following limit (if it exists for
any $\Psi \in {\cal D}$):
\bea
(\delta \otimes \delta ')_{(\alpha,\beta)}(\Psi ) & \equiv & \frac{1}{2}
\lim_{n \rightarrow \infty} \int_{-\infty}^{\infty} [\delta_n(x)^{(\beta)}\,
\delta ' _{red}(x,\frac{1}{n^\alpha}) + \nonumber \\
 & + & \delta _n^{'(\beta)}(x) \,
\delta_{red}(x,\frac{1}{n^\alpha})] \Psi (x) \, dx 
\label{ddp}
\ena

Using the explicit expressions for the quantities in the integral, see
formulas (3.1)-(3.6), we can compute separately the two contributions in
(\ref{ddp}). With the same change of variable as before we get
\beano
I_1\equiv \lim_{n \rightarrow \infty} \int_{-\infty}^{\infty} \delta_n(x)^
{(\beta)}\,
&\delta ' _{red}(x,\frac{1}{n^\alpha})&\Psi (x) \, dx =  \\
 & &\lim_{n
\rightarrow \infty} \frac{-2}{\pi n^{\alpha-3\beta}}\int_{-1}^{1} \frac{t\,
\Phi(t)\, \Psi(t/n^\beta)\, dt} {(t^2+\frac{1}{n^{2(\alpha-\beta)}})^2}
\enano
and 
\beano
I_2\equiv \lim_{n \rightarrow \infty} \int_{-\infty}^{\infty} \delta _n^{'
(\beta)}(x) \, &\delta_{red}(x,\frac{1}{n^\alpha})&\Psi (x) \, dx = \\
& &\lim_{n
\rightarrow \infty} \frac{1}{\pi n^{\alpha-3\beta}} \int_{-1}^{1} \frac{
\Phi_1(t)\, \Psi(t/n^\beta)\, dt} {t^2+\frac{1}{n^{2(\alpha-\beta)}}}
\enano

We have to apply the LDCT to both the contributions above. Again, one can
prove that if  $\alpha-3\beta \geq 0$ both the sequences defining $I_1$ and
$I_2$ are bounded by a (different) positive function which is integrable in
$[-1,1]$. This time we have chosen the form of $\Phi(x)$ like in (\ref{fi})
with $m>2$. These values of $m$ also ensure punctual convergence of the
sequences. 

Like for
the previous example we need to separate two different situations:
$\alpha=3\beta$ and $\alpha>3\beta$. In the first case, using the LDCT it is
very easy to see that $$ I_1 = \frac{-2}{\pi}A_3 \Psi (0) \hspace{2cm} I_2=-I_1.
$$
If $\alpha>3\beta$ we find 
$$
I_1=I_2=0.
$$
Therefore we can conclude that, under the above hypothesis on the delta
sequence, then
\be
(\delta \otimes \delta')_{(\alpha,\beta)}(\Psi)  = 0 \hspace{2cm} \forall
\alpha\geq 3\beta
 \label{res2}
\en

\vspace{4mm}
{\bf Example 3: $(\delta' \otimes \delta')_{(\alpha,\beta)}$}

The product we are interested in is defined, by (\ref{def}):

$$
(\delta' \otimes \delta' )_{(\alpha,\beta)}(\Psi ) \equiv \lim_{n \rightarrow
\infty} \int_{-\infty}^{\infty} \delta_n^{'(\beta)}(x)\,
\delta'_{red}(x,\frac{1}{n^\alpha})\, \Psi (x) \, dx
$$
where $\Psi(x) \in {\cal D}$.

Using (3.2) and (3.5), and performing the change of
variable $t=n^{\beta}x$, we get 
$$ (\delta' \otimes \delta')_{(\alpha,\beta)}(\Psi ) \equiv \lim_{n
\rightarrow \infty} \frac{-2}{\pi n^{\alpha-4\beta}} \int_{-1}^{1}
\frac{t \Phi_1 (t) \Psi (t/n^{\beta}) \, dt}{(t^2+1/n^{2(\alpha-\beta)})^2}
$$

It is immediate to understand that the condition on $\alpha$ and $\beta$ must be
changed and made stronger. In order to satisfy the hypothesis of the LDCT, we
must require now $\alpha-4\beta \geq 0$ and $m>3$. In particular, if we fix
$\alpha=4\beta$ then the sequence $f_n(t)$ converges, punctually, to the
function $f(t)=\frac{-2}{\pi}\frac{\Phi_1(t)}{t^3}\,\Psi(0)$. If
$\alpha>4\beta$ then the limit of $f_n(t)$ is zero. After an integration by
part we conclude that 
\be
    (\delta' \otimes \delta')_{(\alpha,\beta)}(\Psi)  = \left\{
          \begin{array}{ll}
          \frac{-6}{\pi}A_4 \delta(\Psi),    &       \alpha=4\beta \\
           0,   &       \alpha>4\beta. 
       \end{array}
        \right. 
\label{res3}
\en

\vspace{4mm}
{\bf Example 4: $(\delta \otimes \delta'')_{(\alpha,\beta)}$}

From the definition (\ref{def}) we have
\bea
(\delta \otimes \delta'')_{(\alpha,\beta)}(\Psi ) & \equiv & \frac{1}{2}
\lim_{n \rightarrow \infty} \int_{-\infty}^{\infty} [\delta_n(x)^{(\beta)}\,
\delta'' _{red}(x,\frac{1}{n^\alpha}) + \nonumber \\
 & + & \delta _n^{''(\beta)}(x) \,
\delta_{red}(x,\frac{1}{n^\alpha})] \Psi (x) \, dx 
\ena
where $\Psi(x)$ is a function in ${\cal D}$. The two contributions are now
\beano
I_1\equiv \lim_{n \rightarrow \infty} \int_{-\infty}^{\infty} &\delta_n(x)^
{(\beta)}\,&\delta'' _{red}(x,\frac{1}{n^\alpha}) \Psi (x) \, dx =  \\
 & &\lim_{n
\rightarrow \infty} \frac{2}{\pi n^{\alpha-4\beta}}\int_{-1}^{1} \frac{
\Phi(t)\, \Psi(t/n^\beta)\,(3t^2-\frac{1}{n^{2(\alpha-\beta)}}) \, dt}
{(t^2+\frac{1}{n^{2(\alpha-\beta)}})^3} 
\enano
and 
\beano
I_2\equiv \lim_{n \rightarrow \infty} \int_{-\infty}^{\infty} \delta _n^{''
(\beta)}(x) \, &\delta_{red}(x,\frac{1}{n^\alpha})&\Psi (x) \, dx = \\
& &\lim_{n
\rightarrow \infty} \frac{1}{\pi n^{\alpha-4\beta}} \int_{-1}^{1} \frac{
\Phi_2(t)\, \Psi(t/n^\beta)\, dt} {t^2+\frac{1}{n^{2(\alpha-\beta)}}}
\enano

The hypothesis of the LDCT are satisfied if $\Phi(t)$ is like in
(\ref{fi}), with $m>3$ and if $\alpha-4\beta \geq 0$. In particular, if
$\alpha=4\beta$, we get $$
I_1=I_2=\frac{6}{\pi}A_4\, \Psi(0),
$$
while, if $\alpha>4\beta$, 
$$I_1=I_2=0.$$
We can conclude now that
\be
    (\delta \otimes \delta'')_{(\alpha,\beta)}(\Psi)  = \left\{
          \begin{array}{ll}
          \frac{6}{\pi}A_4 \delta(\Psi),    &       \alpha=4\beta \\
           0,   &       \alpha>4\beta. 
       \end{array}
        \right. 
\label{res4}
\en
Incidentally we notice that $(\delta \otimes \delta'')_{(\alpha,\beta)}(\Psi)= -
(\delta' \otimes \delta')_{(\alpha,\beta)}(\Psi)$.

\vspace{4mm}
{\bf Example 5: $(\delta' \otimes \delta'')_{(\alpha,\beta)}$}

We have
\bea
(\delta' \otimes \delta'')_{(\alpha,\beta)}(\Psi ) & \equiv & \frac{1}{2}
\lim_{n \rightarrow \infty} \int_{-\infty}^{\infty} [\delta_n(x)^{'(\beta)}\,
\delta'' _{red}(x,\frac{1}{n^\alpha}) + \nonumber \\
 & + & \delta _n^{''(\beta)}(x) \,
\delta'_{red}(x,\frac{1}{n^\alpha})] \Psi (x) \, dx 
\ena
with $\Psi(x) \in {\cal D}$. The two contributions are now
\beano
I_1\equiv \lim_{n \rightarrow \infty} \int_{-\infty}^{\infty} &\delta_n(x)^
{'(\beta)}\,&\delta'' _{red}(x,\frac{1}{n^\alpha}) \Psi (x) \, dx =  \\
 & &\lim_{n
\rightarrow \infty} \frac{2}{\pi n^{\alpha-5\beta}}\int_{-1}^{1} \frac{
\Phi_1(t)\, \Psi(t/n^\beta)\,(3t^2-\frac{1}{n^{2(\alpha-\beta)}})\, dt}
{(t^2+\frac{1}{n^{2(\alpha-\beta)}})^3} 
\enano
and 
\beano
I_2\equiv \lim_{n \rightarrow \infty} \int_{-\infty}^{\infty} \delta _n^{''
(\beta)}(x) \, &\delta'_{red}(x,\frac{1}{n^\alpha})&\Psi (x) \, dx = \\
& &\lim_{n
\rightarrow \infty} \frac{-2}{\pi n^{\alpha-5\beta}} \int_{-1}^{1} \frac{
\Phi_2(t)\, \Psi(t/n^\beta)\,t \, dt} {(t^2+\frac{1}{n^{2(\alpha-\beta)}})^2}
\enano
In order to satisfy the hypothesis of the LDCT we impose $m>4$ and
$\alpha-5\beta \geq 0$. 

After some manipulation we see that, like for the product between $\delta$ and
$\delta'$, a cancellation occurs between $I_1$ and $I_2$ and therefore 
\be
(\delta' \otimes \delta'')_{(\alpha,\beta)}(\Psi)  = 0 \hspace{2cm} \forall
\alpha\geq 5\beta.
 \label{res5}
\en

\vspace{4mm}
{\bf Example 6: $(\delta'' \otimes \delta'')_{(\alpha,\beta)}$}

The last product we are going to discuss here is defined by the following limit:
$$
(\delta'' \otimes \delta'' )_{(\alpha,\beta)}(\Psi ) \equiv \lim_{n \rightarrow
\infty} \int_{-\infty}^{\infty} \delta_n^{''(\beta)}(x)\,
\delta''_{red}(x,\frac{1}{n^\alpha})\, \Psi (x) \, dx
$$
which can be written as
$$
(\delta'' \otimes \delta'' )_{(\alpha,\beta)}(\Psi ) = \lim_{n \rightarrow
\infty} \frac{2}{\pi n^{\alpha-6\beta}}\int_{-1}^{1} \frac{
\Phi_2(t)\, \Psi(t/n^\beta)\,(3t^2-\frac{1}{n^{2(\alpha-\beta)}}\, dt}
{(t^2+\frac{1}{n^{2(\alpha-\beta)}})^3}
$$
To apply the LDCT we require $m>5$ and $\alpha-6\beta \geq 0$. With
these hypotheses we get
\be
    (\delta'' \otimes \delta'')_{(\alpha,\beta)}(\Psi)  = \left\{
          \begin{array}{ll}
          \frac{120}{\pi}A_6 \delta(\Psi),    &       \alpha=6\beta \\
           0,   &       \alpha>6\beta. 
       \end{array}
        \right. 
\label{res6}
\en

\section{Conclusions}

In this Section we will summarize and comment the results concerning the
previous examples. We will also discuss under which conditions it is possible 
to generalize the results to products of the type $(\delta^{(l)} \otimes
\delta^{(k)} )_ {(\alpha,\beta)}$. We assume all throughout this Section that
the function $\Phi$ has the form in (\ref{fi}).

If $m>1$ we have
\be
    (\delta \otimes \delta)_{(\alpha,\beta)}  = \left\{
          \begin{array}{ll}
          \frac{1}{\pi}A_2 \delta,    &       \alpha=2\beta \\
           0,   &       \alpha>2\beta. 
       \end{array}
        \right. 
\label{resf1}
\en
If $m>2$ then
\be
(\delta \otimes \delta')_{(\alpha,\beta)}  = 0 \hspace{2cm} \forall
\alpha\geq 3\beta
 \label{resf2}
\en
If $m>3$ then
\be
    (\delta' \otimes \delta')_{(\alpha,\beta)}  = \left\{
          \begin{array}{ll}
          \frac{-6}{\pi}A_4 \delta,    &       \alpha=4\beta \\
           0,   &       \alpha>4\beta. 
       \end{array}
        \right. 
\label{resf3}
\en
Again, if $m>3$ then
\be
    (\delta \otimes \delta'')_{(\alpha,\beta)}  = \left\{
          \begin{array}{ll}
          \frac{6}{\pi}A_4 \delta,    &       \alpha=4\beta \\
           0,   &       \alpha>4\beta. 
       \end{array}
        \right. 
\label{resf4}
\en
If $m>4$ then
\be
(\delta' \otimes \delta'')_{(\alpha,\beta)}  = 0 \hspace{2cm} \forall
\alpha\geq 5\beta
 \label{resf5}
\en
And finally, if $m>5$ then
\be
    (\delta'' \otimes \delta'')_{(\alpha,\beta)}  = \left\{
          \begin{array}{ll}
          \frac{120}{\pi}A_6 \delta,    &       \alpha=6\beta \\
           0,   &       \alpha>6\beta. 
       \end{array}
        \right. 
\label{resf6}
\en

The first trivial remark is that, if we choose $m=6$, all the above restrictions
on $m$ are satisfied (and $m$ is even as it is necessary to get the correct
normalization for $\Phi$).

If we choose, for instance, $\alpha=6\beta$, then all the above products are
defined and are all zero but for the last one, $(\delta'' \otimes
\delta'')_{(\alpha,\beta)}$.

We can generalize the above results very easily. If we are interested in
defining a possible product between two distributions like  $\delta^{(l)}$ and
$\delta^{(k)}$, this can be done using (\ref{def}). It is
sufficient to notice that we have to take $m$ (even and) bigger than $l+k+1$
and that a choice $\alpha > (l+k+2)\beta$ gives $(\delta^{(l)} \otimes
\delta^{(k)} )_ {(\alpha,\beta)}=0$, while if $\alpha=(l+k+2)\beta$ then the
product $(\delta^{(l)} \otimes \delta^{(k)} )_ {(\alpha,\beta)}$ can be
different from zero. With this fixed choice of $m, \alpha$ and $\beta$ we also
have  $(\delta^{(i)} \otimes \delta^{(j)} )_ {(\alpha,\beta)}=0$ if $i+j<l+k$.

Before ending this paper it is crucial to observe that the power of definition
(\ref{def}) does not relies on the fact that we fix the function $\Phi$, and
therefore the $\delta$-sequence. If this were so, then we could have done the
same also starting from the sequential completion method. So, let us try to
repeat the same steps as in \cite{col}, Chapter 2, where the non-existence of
the square of a delta function is discussed using sequential completion. If we
use definition (\ref{conv}) and we fix $\Phi$, we still get into the same
troubles as in \cite{col}, because the limit, 
$$ \lim_{n\rightarrow
\infty} n^{\beta} \int_{-1}^{1} \Phi(t)\, \Psi(t/n^{\beta})\, dt,
$$
does not exist for any function $\Psi \in {\cal D}$.
Comparing the above expression with formula (\ref{delta2}) we understand that
a crucial role for the existence of the product is played by $n^{2\beta-
\alpha}$, which converges for any $\alpha \geq 2\beta$. So, the simultaneous
use of the analytic and the sequential completion regularizations of
distributions, seems to be a good idea for extending the product.

We finally want to remark that, even if all throughout this paper the LDCT has
played a crucial role, this is only for technical reasons. One can try to
compute the integral in (\ref{def}) using other techniques, like, for
instance, the one proposed in \cite{li}. In this way one can try to use our
definition to extend the product also to other distributions, even if the limit
in (\ref{def}) cannot be easily computed using LDCT.

\vspace{40pt}

\noindent{\large \bf Appendix: Physical Applications} 

\vspace{5mm}

This Appendix, mainly thought for physicists, shows how definition
(\ref{def}) can be applied to the analysis of some specific quantum mechanical
models. 

We will first discuss a class of models described by hamiltonians
whose eigenstates and eigenvalues can be found using our regularization.

 We will then briefly discuss the meaning of regularization
(\ref{resf1}) in the definition of the probability density (as the square
modulus of the wave function) of a given two-particles system in the classical
limit.

\vspace{5mm}

\noindent
{\bf (a)} We start by considering the following hamiltonian
$$
H_d=-\frac{1}{2}\frac{d^2}{dx^2} +V_o \delta(x)\delta(x-d)
$$
where $d$ is a fixed lenght.

If $d\neq 0$ the hamiltonian becomes the one of a free particle. Its
eigenvectors have the form $
\Psi (x) =A e^{ikx}
$,
with $k^2=2E$, $E$ being the energy of the particle and A a
normalization constant. Notice that this solution does not belong to ${\cal
L}^2(\bf R)$. We will not mind about this, since we can
consider an ${\cal L}^2(\bf R)$ superposition of free particle states with
different $k$, see \cite{mer}.

If $d=0$ the hamiltonian above loses meaning since the product of two delta
functions centered at the same point appears.
 We can consider, in this case, the regularized hamiltonian 
$$
H_{\alpha ,\beta} \equiv -\frac{1}{2}\frac{d^2}{dx^2} +V_o (\delta \otimes \delta)
_{(\alpha,\beta)}.
$$
Choosing $\Phi$ like in (\ref{fi}) with $m$ even and bigger than $1$, we
obtain two different forms of $H_{\alpha ,\beta}$:

\beano
    H_{\alpha,\beta}  = \left\{
          \begin{array}{ll}
          -\frac{1}{2}\frac{d^2}{dx^2} +\frac{V_oA_2}{\pi} \delta (x) ,   
&       \alpha=2\beta \\
           -\frac{1}{2}\frac{d^2}{dx^2},   &       \alpha>2\beta. 
       \end{array}
        \right. 
\enano
The eigensolutions of $H_{2\beta,\beta}$ can be easily found, \cite{haar}:

\beano
    \Psi (x)  = \left\{
          \begin{array}{ll}
          A\{e^{ikx} + \frac{V_oA_2}{i\pi k-V_oA_2} e^{-ikx}\} ,   
&       x<0 \\
          \frac{ikA\pi}{i\pi k-V_oA_2} e^{ikx},   &       x>0. 
       \end{array}
        \right. 
\enano

It is easy to verify that these solutions (still not belonging to $
{\cal L}^2(\bf R)$) are continuous in $x=0$ but their derivatives are not. It
is therefore obvious that any given regularity requirement on the wave
function,
 strongly suggest which regularization to choose, that is the values of
$\alpha$ and $\beta$.

To analogous conclusions we can even arrive for hamiltonians like
$$
H_d=-\frac{1}{2}\frac{d^2}{dx^2} +V_o \delta^{(l)}(x)\delta^{(k)}(x-d)
$$
since all the regularizations found in Section 3 give for
$(\delta^{(i)} \otimes \delta^{(j)} )_ {(\alpha,\beta)}$ only $0$ or a
constant times $\delta (x)$.
\vspace{4mm}

{\bf Remark}--
It may be interesting to notice that a delta function potential has a deep
physical meaning since it is used to describe impurities in solid state
structures. 

\vspace{10mm}
\noindent
{\bf (b)} Let us now consider a two-particles system described by a
factorazible wave function
$$
\Phi^{\epsilon}(x_1,x_2,t)=\Phi^{\epsilon}_1(x_1,t)\Phi^{\epsilon}_2(x_1,t)
$$
where
$$
|\Phi^{\epsilon}_1(x,0)|^2=|\Phi^{\epsilon}_2(x,0)|^2  \equiv
\frac{\exp\{-(x/\epsilon)^2\}}{\epsilon \sqrt{\pi}}.
$$
As it is well known $P_\epsilon (x_1,x_2) \equiv |\Phi^{\epsilon}_1(x_1,0)|^2
|\Phi^{\epsilon}_2(x_2,0)|^2 dx_1 \, dx_2$ is the probability of finding at
$t=0$ particle $1$ between $x_1$ and $x_1+dx_1$ and particle $2$ between $x_2$
and $x_2+dx_2$, \cite{mer}.
In the limit $\epsilon \rightarrow 0$ we get $|\Phi^{\epsilon}_i(x,0)|^2
\rightarrow \delta (x)$ (for instance in ${\cal D'}$), so that
 $P_\epsilon (x_1,x_2)
\rightarrow \delta (x_1)\, \delta (x_2)\, dx_1 \, dx_2$. Because of this we say
that $\epsilon \rightarrow 0$ corresponds to the classical limit of the
system: in fact each particle is sharply centered in a point.

It is possible, therefore, to compute the probability of finding both particles
in the same point $x$, in this classical limit. Of course simple physical
considerations require this probability to be zero. Therefore, since this
probability is proportional to $\delta (x) \delta (x)$, we conclude that the
natural regularization is the one in (\ref{resf1}) with $\alpha >2\beta$.

\vspace{50pt}

\noindent{\large \bf Acknowledgments} \vspace{5mm}

	It is a pleasure to thank Dr. R. Belledonne for her precious help and for her
patience. Thanks are also due to Dr. C. Trapani for his kind reading of the
manuscript.

\end{document}